\documentclass[twocolumn,showpacs,superscriptaddress,amsmath,amssymb,prd]{revtex4-1}
\usepackage[colorlinks,linkcolor=blue,anchorcolor=blue,citecolor=blue,urlcolor=blue]{hyperref}
\usepackage{graphicx}
\usepackage{epsfig}
\usepackage{dcolumn}
\usepackage{bm}
\usepackage{overpic}
\usepackage{color}
\usepackage{xcolor} 
\usepackage{ulem}

\usepackage{multirow}
\usepackage{subfigure}
\usepackage{lineno}
\usepackage{verbatim}
\usepackage{cleveref}
\usepackage{amsthm,amsmath,amssymb}
\usepackage{mathrsfs}
\usepackage{booktabs}
\newcommand\jpsi{J/\psi}
\newcommand\psip{\psi(3686)}

\newcommand{\BESIIIorcid}[1]{\href{https://orcid.org/#1}{\hspace*{0.1em}\raisebox{-0.45ex}{\includegraphics[width=1em]{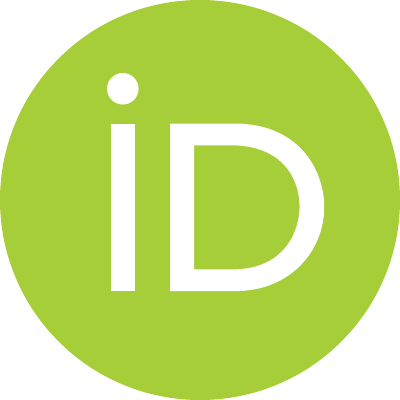}}}}
\begin{document}
\newcommand{\ks}{K_{S}^{0}}
\newcommand{\EP}{e^{+}}
\newcommand{\EM}{e^{-}}
\newcommand{\epm}{e^{\pm}}
\newcommand{\vpho}{\gamma^{\ast}}
\newcommand{\qqbar}{q\bar{q}}

\newcommand{\ee}{e^{+}e^{-}}
\newcommand{\mm}{\mu^{+}\mu^{-}}
\newcommand{\alfs}{\alpha_{s}}
\newcommand{\alfmz}{\alpha(M_{Z}^{2})}
\newcommand{\amu}{a_{\mu}}
\newcommand{\Lam}{\Lambda_{c}}
\newcommand{\lam}{\Lambda_{c}^{+}}
\newcommand{\lambar}{\bar{\Lambda}_{c}^{-}}
\newcommand{\Lambdac}{\Lambda_{c}}
\newcommand{\mbc}{M_{BC}}
\newcommand{\dele}{\Delta E}
\newcommand{\ebm}{E_{\textmd{beam}}}
\newcommand{\ecm}{E_{\textmd{c.m.}}}
\newcommand{\pbm}{p_{\textmd{beam}}}
\newcommand{\MuMu}{\mu\mu}
\newcommand{\mumu}{\mu\mu}
\newcommand{\tata}{\tau^{+}\tau^{-}}
\newcommand{\pipi}{\pi^{+}\pi^{-}}
\newcommand{\gaga}{\gamma\gamma}
\newcommand{\twopho}{\ee+X}
\newcommand{\sqs}{\sqrt{s}}
\newcommand{\sqsp}{\sqrt{s^{\prime}}}
\newcommand{\da}{\Delta\alpha}
\newcommand{\das}{\Delta\alpha(s)}
\newcommand{\dimu}{\ee \ra \mumu}
\newcommand{\dedx}{\textmd{d}E/\textmd{d}x}
\newcommand{\chip}{\chi_{\textmd{Prob}}}
\newcommand{\chiP}{\chi_{p}}
\newcommand{\evz}{V_{z}^{\textmd{evt}}}
\newcommand{\evzloose}{V_{z,\textmd{loose}}^{\textmd{evt}}}
\newcommand{\avz}{V_{z}^{\textmd{ave}}}
\newcommand{\Ngd}{N_{\textmd{good}}}
\newcommand{\Ncru}{N_{\textmd{crude}}}
\newcommand{\pio}{\pi^{0}}
\newcommand{\rpid}{r_{\textmd{PID}}}

\newcommand{\Nhxobs}{N_{h+X}^{\textmd{obs}}}
\newcommand{\Nhobs}{N_{h}^{\textmd{obs}}}
\newcommand{\Npioxobs}{N_{\pi^{0}+X}^{\textmd{obs}}}
\newcommand{\Nksxobs}{N_{\ks+X}^{\textmd{obs}}}
\newcommand{\Npioobs}{N_{\pi^{0}}^{\textmd{obs}}}
\newcommand{\Nksobs}{N_{\ks}^{\textmd{obs}}}
\newcommand{\Nhxtru}{N_{h+X}^{\textmd{tru}}}
\newcommand{\Nhtru}{N_{h}^{\textmd{tru}}}
\newcommand{\Npiotru}{N_{\pi^{0}}^{\textmd{tru}}}
\newcommand{\Nkstru}{N_{\ks}^{\textmd{tru}}}
\newcommand{\Nbarhxobs}{\bar{N}_{h+X}^{\textmd{obs}}}
\newcommand{\Nbarhobs}{\bar{N}_{h}^{\textmd{obs}}}
\newcommand{\Nbarpioobs}{\bar{N}_{\pi^{0}}^{\textmd{obs}}}
\newcommand{\Nbarksobs}{\bar{N}_{\ks}^{\textmd{obs}}}
\newcommand{\Nbarhxtru}{\bar{N}_{h+X}^{\textmd{tru}}}
\newcommand{\Nbarhtru}{\bar{N}_{h}^{\textmd{tru}}}
\newcommand{\Nbarpiotru}{\bar{N}_{\pi^{0}}^{\textmd{tru}}}
\newcommand{\Nbarkstru}{\bar{N}_{\ks}^{\textmd{tru}}}

\newcommand{\Nhadtot}{N_{\textmd{had}}^{\textmd{tot}}}
\newcommand{\Nhadobs}{N_{\textmd{had}}^{\textmd{obs}}}
\newcommand{\Nbarhadobs}{\bar{N}_{\textmd{had}}^{\textmd{obs}}}
\newcommand{\Nhadtru}{N_{\textmd{had}}^{\textmd{tru}}}
\newcommand{\Nbarhadtru}{\bar{N}_{\textmd{had}}^{\textmd{tru}}}
\newcommand{\Nhadphy}{N_{\textmd{had}}}

\newcommand{\cshadobs}{\sigma_{\textmd{had}}^{\textmd{obs}}}
\newcommand{\effhad}{\vap_{\textmd{had}}}
\newcommand{\efftrg}{\vap_{\textmd{trig}}}
\newcommand{\lint}{\mathcal{L}_{\textmd{int}}}
\newcommand{\Nbkg}{N_{\textmd{bkg}}}
\newcommand{\NbkgTot}{N_{\textrm{bkg}}^{\textrm{Tot}}}
\newcommand{\csbkg}{\sigma_{\textmd{bkg}}}
\newcommand{\Nmcsur}{N_{\textmd{MC}}^{\textmd{sur}}}
\newcommand{\Nmcsurori}{N_{\textmd{MC}}^{\textmd{sur,nom.}}}
\newcommand{\Nmcsurwtd}{N_{\textmd{MC}}^{\textmd{sur,wtd.}}}
\newcommand{\Nmcgen}{N_{\textmd{MC}}^{\textmd{gen}}}
\newcommand{\vap}{\varepsilon}
\newcommand{\chisq}{\chi^{2}}
\newcommand{\cshadphy}{\sigma_{\textmd{had}}^{\textmd{phy}}}
\newcommand{\cshadtot}{\sigma_{\textmd{had}}^{\textmd{tot}}}
\newcommand{\cshadborn}{\sigma_{\textmd{had}}^{0}}
\newcommand{\cshadborncon}{\sigma_{\textmd{con}}^{0}}
\newcommand{\cshadbornres}{\sigma_{\textmd{res}}^{0}}
\newcommand{\csdimuborn}{\sigma_{\mu\mu}^{0}}
\newcommand{\rpqcd}{R_{\textmd{pQCD}}}
\newcommand{\Nprod}{N_{\textmd{prod}}}
\newcommand{\Nhadnet}{N_{\textrm{had}}^{\textrm{net}}}
\newcommand{\Delrel}{\Delta_{\textrm{rel}}}

\newcommand{\fourpionchg}{\pipi\pipi}
\newcommand{\fourpionneu}{\pipi\pi^{0}\pi^{0}}
\newcommand{\sixpionchg}{3(\pipi)}
\newcommand{\thrpionneu}{\pipi\pi^{0}}
\newcommand{\twopionchg}{\pipi}

\newcommand{\Nsurnpion}{N_{\textmd{sur}}^{n\pi}}
\newcommand{\Ngennpion}{N_{\textmd{gen}}^{n\pi}}
\newcommand{\Ngentot}{N_{\textmd{gen}}^{\textmd{tot}}}
\newcommand{\effincnpion}{\vap_{n\pi}^{\textmd{inc}}}
\newcommand{\effincnpionp}{\vap_{n\pi}^{\textmd{inc},\prime}}
\newcommand{\effincnonnpion}{\vap_{\textmd{non}-n\pi}^{\textmd{inc}}}
\newcommand{\effexcnpion}{\vap_{n\pi}^{\textmd{exc}}}
\newcommand{\fracnpion}{f_{n\pi}}
\newcommand{\fracnpionp}{f_{n\pi}^{\prime}}
\newcommand{\fracnonnpion}{f_{\textmd{non}-n\pi}}

\newcommand{\Nsurtwopion}{N_{\textmd{sur}}^{2\pi}}
\newcommand{\Ngentwopion}{N_{\textmd{gen}}^{2\pi}}
\newcommand{\effinctwopion}{\vap_{2\pi}^{\textmd{inc}}}
\newcommand{\effinctwopionp}{\vap_{2\pi}^{\textmd{inc},\prime}}
\newcommand{\effincnontwopion}{\vap_{\textmd{non}-2\pi}^{\textmd{inc}}}
\newcommand{\effexctwopion}{\vap_{2\pi}^{\textmd{exc}}}
\newcommand{\fractwopion}{f_{2\pi}}
\newcommand{\fractwopionp}{f_{2\pi}^{\prime}}
\newcommand{\fracnontwopion}{f_{\textmd{non}-2\pi}}

\newcommand{\Nsurthrpion}{N_{\textmd{sur}}^{3\pi}}
\newcommand{\Ngenthrpion}{N_{\textmd{gen}}^{3\pi}}
\newcommand{\effincthrpion}{\vap_{3\pi}^{\textmd{inc}}}
\newcommand{\effincthrpionp}{\vap_{3\pi}^{\textmd{inc},\prime}}
\newcommand{\effincnonthrpion}{\vap_{\textmd{non}-3\pi}^{\textmd{inc}}}
\newcommand{\effexcthrpion}{\vap_{3\pi}^{\textmd{exc}}}
\newcommand{\fracthrpion}{f_{3\pi}}
\newcommand{\fracthrpionp}{f_{3\pi}^{\prime}}
\newcommand{\fracnonthrpion}{f_{\textmd{non}-3\pi}}

\newcommand{\Nsurfourpion}{N_{\textmd{sur}}^{4\pi}}
\newcommand{\Ngenfourpion}{N_{\textmd{gen}}^{4\pi}}
\newcommand{\effincfourpion}{\vap_{4\pi}^{\textmd{inc}}}
\newcommand{\effincfourpionp}{\vap_{4\pi}^{\textmd{inc},\prime}}
\newcommand{\effincnonfourpion}{\vap_{\textmd{non}-4\pi}^{\textmd{inc}}}
\newcommand{\effexcfourpion}{\vap_{4\pi}^{\textmd{exc}}}
\newcommand{\fracfourpion}{f_{4\pi}}
\newcommand{\fracfourpionp}{f_{4\pi}^{\prime}}
\newcommand{\fracnonfourpion}{f_{\textmd{non}-4\pi}}

\newcommand{\Npionprod}{N_{\textmd{prod}}^{4\pi}}
\newcommand{\Ndatasur}{N_{\textmd{data}}^{\textmd{sur}}}
\newcommand{\Nobspion}{N_{\textmd{obs}}^{4\pi}}
\newcommand{\Nhadprod}{N_{\textmd{prod}}^{\textmd{had}}}
\newcommand{\sigmaobs}{\sigma_{\textmd{obs}}}
\newcommand{\effhadp}{\vap_{\textmd{had}}^{\prime}}

\newcommand{\effpion}{\vap_{4\pi}}
\newcommand{\effexcpion}{\vap_{4\pi}^{\textmd{exc}}}
\newcommand{\effincpion}{\vap_{4\pi}^{\textmd{inc}}}
\newcommand{\effincpionI}{\vap_{4\pi}^{\textmd{inc},1}}
\newcommand{\effincpionII}{\vap_{4\pi}^{\textmd{inc},2}}
\newcommand{\effincpionp}{\vap_{4\pi}^{\textmd{inc},\prime}}
\newcommand{\effincremain}{\vap_{\textmd{non}-n\pi}^{\textmd{inc}}}

\newcommand{\fracpion}{f_{4\pi}}
\newcommand{\fracnonpion}{f_{\textmd{non}-4\pi}}
\newcommand{\fracnonpionp}{f_{\textmd{non}-4\pi}^{\prime}}
\newcommand{\fracpionII}{f_{4\pi}^{2}}
\newcommand{\fracpionp}{f_{4\pi}^{\prime}}
\newcommand{\reladiff}{\Delta_{\textmd{rel}}}

\newcommand{\Nsursixpion}{N_{\textmd{sur}}^{6\pi}}
\newcommand{\Ngensixpion}{N_{\textmd{gen}}^{6\pi}}
\newcommand{\effincsixpion}{\vap_{6\pi}^{\textmd{inc}}}
\newcommand{\fracsixpion}{f_{6\pi}}

\newcommand{\etot}{E_{\textmd{tot}}}
\newcommand{\ptot}{p_{\textmd{tot}}}
\newcommand{\plab}{p_{\textmd{Lab}}}
\newcommand{\mpiOI}{M(\pi^{0}_{1})}
\newcommand{\mpiOII}{M(\pi^{0}_{2})}

\newcommand{\widtheeoi}{\varGamma^{\textmd{ee}}_{0,i}}
\newcommand{\widtheeoj}{\varGamma^{\textmd{ee}}_{0,j}}
\newcommand{\widtheeo}{\varGamma^{\textmd{ee}}_{0}}
\newcommand{\widthee}{\varGamma^{\textmd{ee}}}
\newcommand{\widtheeexpi}{\varGamma^{\textmd{ee}}_{\textmd{exp},i}}
\newcommand{\widtheeexp}{\varGamma^{\textmd{ee}}_{\textmd{exp}}}
\newcommand{\widthtoti}{\varGamma^{\textmd{tot}}_{i}}
\newcommand{\widthtot}{\varGamma^{\textmd{tot}}}

\newcommand{\vpqed}{\Pi_{\textmd{QED}}}
\newcommand{\vpqcd}{\Pi_{\textmd{QCD}}}
\newcommand{\vpcon}{\Pi_{\textmd{con}}}
\newcommand{\vpres}{\Pi_{\textmd{res}}}
\newcommand{\vpo}{\Pi_{0}}
\newcommand{\rcon}{R_{\textmd{con}}}
\newcommand{\rres}{R_{\textmd{res}}}
\newcommand{\rexp}{R_{\textmd{exp}}}

\newcommand{\delvert}{\delta_{\textmd{vert}}}
\newcommand{\delvp}{\delta_{\textmd{vac}}}
\newcommand{\delbrem}{\delta_{\gamma}}
\newcommand{\delobs}{\delta_{\textmd{obs}}}
\newcommand{\radiatorsf}{F_{\textmd{SF}}}
\newcommand{\radiatorfd}{F_{\textmd{FD}}}
\newcommand{\DelFD}{\Delta_{\textmd{FD}}}
\newcommand{\DelFDCal}{\Delta_{\textmd{cal}}}
\newcommand{\DelFDcs}{\Delta_{\sigma}}
\newcommand{\DelFDvp}{\Delta_{\textmd{vp}}}

\newcommand{\costh}{\cos\theta}
\newcommand{\costhIprg}{\cos\theta_{\textmd{1prg}}}
\newcommand{\costhIIprg}{\cos\theta_{\textmd{2prg}}}
\newcommand{\costhIIIprg}{\cos\theta_{\textmd{3prg}}}
\newcommand{\costhIVprg}{\cos\theta_{\textmd{4prg}}}
\newcommand{\costhrestprg}{\cos\theta_{\textmd{restprg}}}
\newcommand{\emce}{E^{\textmd{ctrk.}}_{\textmd{emc}}}
\newcommand{\emceIprg}{E^{\textmd{ctrk.}}_{\textmd{emc,1prg}}}
\newcommand{\emceIIprg}{E^{\textmd{ctrk.}}_{\textmd{emc,2prg}}}
\newcommand{\emceIIIprg}{E^{\textmd{ctrk.}}_{\textmd{emc,3prg}}}
\newcommand{\emceIVprg}{E^{\textmd{ctrk.}}_{\textmd{emc,4prg}}}
\newcommand{\emcerestprg}{E^{\textmd{ctrk.}}_{\textmd{emc,restprg}}}
\newcommand{\isocosth}{\cos\theta_{\textmd{iso}}}
\newcommand{\isocosthIprg}{\cos\theta_{\textmd{iso,1prg}}}
\newcommand{\isocosthIIprg}{\cos\theta_{\textmd{iso,2prg}}}
\newcommand{\isocosthIIIprg}{\cos\theta_{\textmd{iso,3prg}}}
\newcommand{\isocosthIVprg}{\cos\theta_{\textmd{iso,4prg}}}
\newcommand{\isocosthrestprg}{\cos\theta_{\textmd{iso,restprg}}}
\newcommand{\eop}{E/P}
\newcommand{\eopIprg}{E/P_{\textmd{1prg}}}
\newcommand{\eopIIprg}{E/P_{\textmd{2prg}}}
\newcommand{\eopIIIprg}{E/P_{\textmd{3prg}}}
\newcommand{\eopIVprg}{E/P_{\textmd{4prg}}}
\newcommand{\eoprestprg}{E/P_{\textmd{restprg}}}
\newcommand{\nisogam}{N_{\textmd{isogam}}}
\newcommand{\nisogamIprg}{N_{\textmd{isogam,1prg}}}
\newcommand{\nisogamIIprg}{N_{\textmd{isogam,2prg}}}
\newcommand{\nisogamIIIprg}{N_{\textmd{isogam,3prg}}}
\newcommand{\nisogamIVprg}{N_{\textmd{isogam,4prg}}}
\newcommand{\nisogamrestprg}{N_{\textmd{isogam,restprg}}}
\newcommand{\ptrk}{p_{\textmd{ctrk}}}
\newcommand{\pIprg}{p^{\textmd{ctrk}}_{\textmd{1prg}}}
\newcommand{\pIIprg}{p^{\textmd{ctrk}}_{\textmd{2prg}}}
\newcommand{\pIIIprg}{p^{\textmd{ctrk}}_{\textmd{3prg}}}
\newcommand{\pIVprg}{p^{\textmd{ctrk}}_{\textmd{4prg}}}
\newcommand{\prestprg}{p^{\textmd{ctrk}}_{\textmd{restprg}}}
\newcommand{\tote}{E_{\textmd{vis.}}}
\newcommand{\totevte}{E_{\textmd{tot.}}}
\newcommand{\totevteIprg}{E_{\textmd{tot.}}^{\textmd{1prg}}}
\newcommand{\balanceIprg}{\textmd{Balance}}
\newcommand{\ngamma}{N_{\gamma}}
\newcommand{\ngammaIprg}{N_{\gamma,\textmd{1prg}}}
\newcommand{\ngammaIIprg}{N_{\gamma,\textmd{2prg}}}
\newcommand{\ngammaIIIprg}{N_{\gamma,\textmd{3prg}}}
\newcommand{\ngammaIVprg}{N_{\gamma,\textmd{4prg}}}
\newcommand{\ngammarestprg}{N_{\gamma,\textmd{restprg}}}
\newcommand{\ngoodwt}{N_{\textmd{good}}^{\textmd{Wt}}}
\newcommand{\ngood}{N_{\textmd{good}}}
\newcommand{\npiO}{N_{\pi^{0}}}
\newcommand{\npiOIprg}{N_{\pi^{0}}^{\textmd{1prg}}}
\newcommand{\npiOIIprg}{N_{\pi^{0}}^{\textmd{2prg}}}
\newcommand{\npp}{N_{p}}
\newcommand{\npm}{N_{\bar{p}}}
\newcommand{\nkp}{N_{K^{+}}}
\newcommand{\nkm}{N_{K^{-}}}
\newcommand{\npip}{N_{\pi^{+}}}
\newcommand{\npim}{N_{\pi^{-}}}
\newcommand{\ppp}{P(p^{+})}
\newcommand{\ppm}{P(\bar{p}^{-})}
\newcommand{\pkp}{p(K^{+})}
\newcommand{\pkm}{p(K^{-})}
\newcommand{\ppip}{P(\pi^{+})}
\newcommand{\ppim}{P(\pi^{-})}
\newcommand{\ppiO}{P(\pi^{0})}
\newcommand{\mpiO}{M(\pi^{0})}
\newcommand{\mks}{M(K^{0}_{s})}
\newcommand{\pks}{p_{K^{0}_{s}}}
\newcommand{\mphi}{M(\phi)}
\newcommand{\pphi}{p_{\phi}}
\newcommand{\mIIgam}{M(\gamma\gamma)}
\newcommand{\mIIgamIprg}{M(\gamma\gamma)^{\textmd{1prg}}}
\newcommand{\pIIgam}{p_{\gamma\gamma}}
\newcommand{\mlambda}{M(\Lambda)}
\newcommand{\plambda}{p_{\Lambda}}
\newcommand{\mdO}{M(D^{0})}
\newcommand{\pdO}{p_{D^{0}}}
\newcommand{\mdstarO}{M(D^{\ast 0})}
\newcommand{\pdstarO}{p_{D^{\ast 0}}}
\newcommand{\mdp}{M(D^{\pm})}
\newcommand{\pdp}{p_{D^{\pm}}}
\newcommand{\mdstarp}{M(D^{\ast\pm})}
\newcommand{\pdstarp}{p_{D^{\ast\pm}}}
\newcommand{\mds}{M(D_{s}^{\pm})}
\newcommand{\pds}{p_{D_{s}^{\pm}}}
\newcommand{\mdstars}{M(D_{s}^{\ast\pm})}
\newcommand{\pdstars}{p_{D_{s}^{\ast\pm}}}
\newcommand{\Vr}{V_{r}}
\newcommand{\Vz}{V_{z}}

\newcommand{\gev}{\mathrm{GeV}}
\newcommand{\mev}{\mathrm{MeV}}
\newcommand{\mevcc}{\mathrm{MeV}/c^{2}}
\newcommand{\gevc}{\mathrm{GeV}/c}
\newcommand{\gevcc}{\mathrm{GeV}/c^2}

\newcommand{\nchg}{N_{\textmd{chg}}}
\newcommand{\eff}{\vap}

\newcommand{\critecm}{1.780}

\newcommand{\ENERGYAT}{4575.5}
\newcommand{\ENERGYBT}{4575.5}
\newcommand{\ENERGYCT}{4575.5}
\newcommand{\ENERGYDT}{4575.5}
\newcommand{\ksdecay}{\ks\ra\pi^{+}\pi^{-}}
\newcommand{\phidecay}{\phi\ra K^{+}K^{-}}
\newcommand{\piOdecay}{\pi^{0}\ra\gamma\gamma}
\newcommand{\Lambdadecay}{\Lambda\ra p\pi^{-}}
\newcommand{\DOdecay}{D^{0}\ra K^{-}\pi^{+}}
\newcommand{\DStarOdecay}{D^{\ast0}\ra D^{0}\pi^{0}}
\newcommand{\Dpdecay}{D^{+}\ra K^{+}\pi^{+}\pi^{-}}
\newcommand{\DStarpdecay}{D^{\ast+}\ra D^{0}\pi^{+}}
\newcommand{\Dsdecay}{D^{+}_{s}\ra K^{+}K^{-}\pi^{+}}
\newcommand{\DStarsdecay}{D^{\ast+}_{s}\ra D^{+}_{s}\gamma}


\title{\boldmath\textbf{ \textbf{First measurements}} of the branching fractions of $J/\psi$ and $\psip \to \Sigma^{0} \bar{\Sigma}^{0}\eta$} 
\author{
M.~Ablikim$^{1}$\BESIIIorcid{0000-0002-3935-619X},
M.~N.~Achasov$^{4,c}$\BESIIIorcid{0000-0002-9400-8622},
P.~Adlarson$^{83}$\BESIIIorcid{0000-0001-6280-3851},
X.~C.~Ai$^{89}$\BESIIIorcid{0000-0003-3856-2415},
C.~S.~Akondi$^{31A,31B}$\BESIIIorcid{0000-0001-6303-5217},
R.~Aliberti$^{39}$\BESIIIorcid{0000-0003-3500-4012},
A.~Amoroso$^{82A,82C}$\BESIIIorcid{0000-0002-3095-8610},
Q.~An$^{79,65,\dagger}$,
Y.~H.~An$^{89}$\BESIIIorcid{0009-0008-3419-0849},
Y.~Bai$^{63}$\BESIIIorcid{0000-0001-6593-5665}
% ...
\\
\vspace{0.2cm}
(BESIII Collaboration)\\
\vspace{0.2cm} {\it
$^{1}$ Institute of High Energy Physics, Beijing 100049, People's Republic of China\\
% ...
$^{4}$ Budker Institute of Nuclear Physics SB RAS (BINP), Novosibirsk 630090, Russia\\
% ...
$^{31}$ INFN Sezione di Ferrara, (A)INFN Sezione di Ferrara, I-44122, Ferrara, Italy; (B)University of Ferrara, I-44122, Ferrara, Italy\\
% ...
$^{39}$ Johannes Gutenberg University of Mainz, Johann-Joachim-Becher-Weg 45, D-55099 Mainz, Germany\\
% ...
$^{63}$ Southeast University, Nanjing 211100, People's Republic of China\\
% ...
$^{65}$ State Key Laboratory of Particle Detection and Electronics, Beijing 100049, Hefei 230026, People's Republic of China\\
% ...
$^{79}$ University of Science and Technology of China, Hefei 230026, People's Republic of China\\
% ...
$^{82}$ University of Turin and INFN, (A)University of Turin, I-10125, Turin, Italy; (B)University of Eastern Piedmont, I-15121, Alessandria, Italy; (C)INFN, I-10125, Turin, Italy\\
$^{83}$ Uppsala University, Box 516, SE-75120 Uppsala, Sweden\\
% ...
$^{89}$ Zhengzhou University, Zhengzhou 450001, People's Republic of China\\

\vspace{0.2cm}
$^{\dagger}$ Deceased\\
% ...
$^{c}$ Also at the Novosibirsk State University, Novosibirsk, 630090, Russia\\
% ...
}

}
\date{\today}

\begin{abstract}

Based on $(10087 \pm 44) \times 10^6$ $J/\psi$ and $(2712 \pm 14) \times 10^6$  $\psi(3686)$ events collected with the BESIII detector at the BEPCII collider, the hadronic decays $J/\psi \to \Sigma^{0} \bar{\Sigma}^{0} \eta$ and $\psi(3686) \to \Sigma^{0} \bar{\Sigma}^{0} \eta$ are observed for the first time.~The corresponding branching fractions are measured to be $\mathcal{B}(J/\psi \to \Sigma^{0} \bar{\Sigma}^{0}\eta)= (7.5 \pm 0.3 \pm 0.8) \times 10^{-5}$ and $\mathcal{B}(\psi(3686) \to \Sigma^{0} \bar{\Sigma}^{0}\eta)= (1.3\pm 0.1 \pm 0.1) \times 10^{-5}$, respectively, where the first uncertainties are statistical, and the second systematic.~The ratio $\text{Q} \approx \frac{\mathcal{B}(\psi(3686) \to \Sigma^{0} \bar{\Sigma}^{0} \eta)}{\mathcal{B}(J/\psi \to \Sigma^{0} \bar{\Sigma}^{0} \eta)}$ is determined to be $(17.3 \pm 1.5 \pm 1.7)\%$, which is consistent with the 12\%-rule within 3.0$\sigma$.~No significant intermediate states or threshold enhancements are observed in the $\Sigma^0$($\bar{\Sigma}^{0}$)$\eta$ and $\Sigma^0$$\bar{\Sigma}^{0}$ invariant mass spectra.

\end{abstract}

\maketitle
\section{Introduction}
Studies of the hadronic decays of the charmonium states $\jpsi$ and $\psip$ provide good opportunities to test quantum chromodynamics (QCD) in the transition region between perturbative and non-perturbative regimes \cite{ref_ J. Beringer-2012}. Within perturbative QCD, the ratio of the branching fractions for $\jpsi$ and $\psip$ decaying into the same final state is predicted to follow the so-called “12$\%$-rule” \cite{ref_1211,ref_1212,ref_1213,ref_1214}, expressed as 
\begin{equation}
Q_h = \frac{\mathcal{B}(\psi(3686) \to h)}{\mathcal{B}(J/\psi \to h)} \simeq \frac{\mathcal{B}(\psi(3686) \to e^+ e^-)}{\mathcal{B}(J/\psi \to e^+ e^-)} \simeq 12\%,
\end{equation}
where $h$ denotes any exclusive hadronic decay mode. Many measured decays are consistent with this prediction within uncertainties.~However, this rule was first observed to be violated in the process $\psi(2S)\to\rho\pi$, which is known as the “$\rho\pi$ puzzle”~\cite{ref_121,ref_llp}.~Subsequent experimental studies have revealed further deviations from the 12$\%$ ratio, such as in the isospin-violating decay $\jpsi$ and $\psi(3686) \to \Lambda\bar{\Lambda}\pi^0$, in contrast to the isospin-conserving decay $\jpsi$ and $\psi(3686) \to \Lambda\bar{\Lambda}\eta$~\cite{ref_M. Ablikim2013}. Further measurements of baryonic final states are essential for improving our understanding of the origin of these discrepancies.

Many fundamental issues in baryon spectroscopy remain poorly understood. Charmonium decays offer an excellent platform for studying excited nucleons and hyperons - $N^*$, $\Lambda^*$, $ \Sigma^*$ and $\Xi^*$ \cite{ref_B. -S. Zou2001}.~Specifically, three-body decays like $B\bar{B}P$, where $B$ denotes a baryon and $P$ represents a pseudoscalar meson ($\pi$,~$\eta$), provide a powerful avenue to search for excited baryon states.~Enhancements near the baryon-antibaryon threshold have been observed in charmonium and $B$ meson decays \cite{ref_Y.-J. Lee,ref_M. Ablikim}.~Numerous theoretical proposals, including gluonic mechanisms, fragmentation mechanisms, baryonia, multiquark states, and final-state interactions, have been proposed to explain these structures \cite{ref_J.Z. Bai2003}.~More searches for such enhancements across different quantum numbers and production channels are valuable for distinguishing between these models.~The BESIII experiment has collected the world’s largest $\jpsi$ and $\psi(3686)$ datasets \cite{ref_psinumber,ref_jpsinumber}, thereby providing a unique opportunity to explore those states.

In this work, we report the first observations and branching fraction measurements of  $J/\psi \to \Sigma^{0} \bar{\Sigma}^{0}\eta$ and $\psi(3686)\to \Sigma^{0} \bar{\Sigma}^{0}\eta$ using $(10087 \pm 44) \times 10^6$ $J/\psi$ and $(2712 \pm 14) \times 10^6$ $\psi(3686)$ events collected with the BESIII detector at the BEPCII collider. Based on these measurements, we test the 12$\%$-rule in the $J/\psi \to \Sigma^{0} \bar{\Sigma}^{0}\eta$ and $\psi(3686) \to \Sigma^{0} \bar{\Sigma}^{0}\eta$ decays. Potential intermediate states or threshold enhancements in the $\Sigma^{0} \bar\Sigma^{0}\eta$ and $\Sigma^{0}\bar\Sigma^{0}$ mass spectra are also examined.

\section{BESIII DETECTOR AND MONTE CARLO SIMULATION}
The BESIII detector~\cite{Ablikim:2009aa} records symmetric $e^+e^-$ collisions 
provided by the BEPCII storage ring~\cite{Yu:IPAC2016-TUYA01} in the center-of-mass energy range from 1.84 to 4.95~GeV, with a peak luminosity of $1.1 \times 10^{33}\;\text{cm}^{-2}\text{s}^{-1}$ achieved at $\sqrt{s} = 3.773\;\text{GeV}$.~BESIII has collected large data samples in this energy region~\cite{Ablikim:2019hff, EcmsMea, EventFilter}. The cylindrical core of the BESIII detector covers 93\% of the full solid angle and consists of a helium-based multilayer drift chamber~(MDC), a time-of-flight system~(TOF), and a CsI(Tl) electromagnetic calorimeter~(EMC), which are all enclosed in a superconducting solenoidal magnet providing a 1.0~T magnetic field. The magnetic field was 0.9~T in 2012. The solenoid is supported by an octagonal flux-return yoke with resistive plate counter muon identification modules interleaved with steel.~The charged-particle momentum resolution at $1~{\rm GeV}/c$ is $0.5\%$, and the 
${\rm d}E/{\rm d}x$ resolution is 6$\%$ for electrons from Bhabha scattering. The EMC measures photon energies with a resolution of $2.5\%$ ($5\%$) at $1$~GeV in the barrel (end cap) region.~The time resolution in the plastic scintillator TOF barrel region is 68~ps, while that in the end cap region was 110~ps.~The end cap TOF system was upgraded in 2015 using multigap resistive plate chamber technology, achieving a time resolution of 60~ps.~This upgrade benefits 87$\%$ of the $J/\psi$ data sample and 83$\%$ of the $\psi(3686)$ data sample used in this analysis, enhancing particle identification for these datasets \cite{60,61,70}.

This analysis is performed using BESIII Offline Software System~\cite{ref_13}.~To investigate potential background and optimize event selection criteria, samples of inclusive $\jpsi$ and $\psi(3686)$ Monte Carlo (MC) events are produced.~The production of the $\jpsi$ or $\psi(3686)$ resonance is simulated by the MC event generator KKMC \cite{ref_KKMC1} including the beam energy spread, and the decays are generated by EVTGEN \cite{ref_Evtgen1, ref_Evtgen2} for known decay modes with branching fractions being set to the Particle Data Group (PDG) world average values~\cite{pdg1}, and by  \textsc{LUNDCHARM}~\cite{ref_Lundcharm} for the remaining unknown decays.~To evaluate the detection efficiencies and optimize the event selection criteria, samples of 0.2 million phase space (PHSP) MC events \cite{ref_29}  are generated for each data-taking year, taking into account the corresponding detector conditions.

\section{event selection}\label{eventsec}
In the channel $\psi(3686) \to \Sigma^{0} \bar{\Sigma}^{0}\eta$, the $\Sigma^{0}$ and $\bar{\Sigma}^{0}$ baryons are reconstructed via the $\gamma \Lambda$ and $\gamma \bar{\Lambda}$ decays, respectively.~The $\eta$ meson is reconstructed via the $\eta\to\gamma\gamma$ decay mode, while the $\Lambda$ and $\bar{\Lambda}$ are reconstructed via the $p \pi^-$ and $\bar{p} \pi^+$ modes, respectively.

Charged tracks detected in the MDC are required to be within a polar angle ($\theta$) range of $|\rm{cos\theta}|<0.93$, where $\theta$ is defined with respect to the detector symmetry axis ($z$-axis), and their distance of closest approach to the interaction point must be less than 20\,cm along the $z$-axis ($R_z$) and less than 10\,cm in the transverse plane ($R_{xy}$).~Particle identification (PID) for charged tracks combines measurements of the energy deposited in the MDC (${\rm d}E/{\rm d}x$) and the flight time in the TOF to form likelihoods $\mathcal{L}(h)$ ($h=p,K,\pi$) for each hadron $h$ hypothesis.~Each track is assigned as the particle type with the greatest PID likelihood.~At least two oppositely charged protons and pions are required in each event. Photon candidates are identified using isolated showers in the EMC. The deposited energy of each shower must be more than 25 MeV in the barrel region $(|\cos{\theta}| < 0.80)$ and more than 50 MeV in the end cap region $(0.86 < |\cos\theta| < 0.92)$. To suppress electronic noise and showers unrelated to the event, the difference between the EMC time and the event start time is required to be within [0, 700] ns. After the above track-level requirements, the candidate events with at least four good photons are selected.

The $\Lambda(\bar{\Lambda})$ candidates are reconstructed by combining identified (anti)protons with oppositely charged pion tracks and performing a secondary vertex fit.~The candidate~with~the~minimum $\chi^2_{\Lambda}$$(\chi^2_{\bar{\Lambda}})$ value is selected.~After reconstructing the $\Lambda$ and $\bar{\Lambda}$ candidates, no additional charged  track within $R_{xy} <1.0$~cm and $R_z <10.0$~cm remains.

To improve momentum resolution and suppress background, a four-constraint (4C) kinematic fit enforcing energy-momentum conservation is performed under the hypothesis of $ \Lambda \bar{\Lambda}\gamma\gamma\gamma\gamma$. The fit constrains the total four-momentum of the final-state particles to that of the initial $e^+e^-$ system.~For events with more than four photons, the $\Lambda \bar{\Lambda}\gamma\gamma\gamma\gamma$ combination with the minimum $\chi_{4\rm{C}}^2$ is retained.~Additionally, a requirement of $\chi_{4\rm{C}}^2<60$ is imposed for $J/\psi \to \Sigma^{0} \bar{\Sigma}^{0}\eta$, while $\chi_{4\rm{C}}^2<40$ is required for $\psi(3686) \to \Sigma^{0} \bar{\Sigma}^{0}\eta$.~The $\chi_{4\rm{C}}^2$ requirements are optimized to maximize the figure of merit defined as $S/\sqrt{S+B}$, where $S$ is the signal yield and $B$ is the background yield from the inclusive MC sample.~The $\Sigma^{0}$, $\bar{\Sigma}^0$ and $\eta$ candidates are selected by minimizing $\chi^2_{\Sigma^{0}\bar{\Sigma}^{0}\eta}\equiv \frac{(M_{\gamma_{1}\gamma_{2}} - M_{\eta})^{2}}{\sigma_{\eta}^{2}} + \frac{(M_{\gamma_{3}\Lambda} - M_{\Sigma^0})^{2}}{\sigma_{\Sigma^0}^{2}} + \frac{(M_{\gamma_{4}\bar{\Lambda}} - M_{\bar{\Sigma}^{0}})^{2}}{\sigma_{\bar{\Sigma}^{0}}^{2}}$, where $M_\eta$, $M_{\Sigma^0}$ and $M_{\bar{\Sigma}^{0}}$ are the masses of $\eta$ and $\Sigma$ cited from the PDG~\cite{pdg1}, respectively, $\sigma_\eta$ and $\sigma_\Sigma$ are the mass resolutions of $\eta$ and $\Sigma$ determined from signal MC samples.~The $\Sigma^{0}$, $\bar{\Sigma}^{0}$ and $\eta$ candidates are required to satisfy $|M_{\gamma \Lambda} - m_{\Sigma^{0}}| < 15$ MeV/$c^{2}$, $|M_{\gamma \bar{\Lambda}} - m_{\bar{\Sigma}^{0}}| < 15$ MeV/$c^{2}$ and $|M_{\gamma \gamma}- m_{\eta}| < 25$ MeV/$c^{2}$, respectively.

To suppress background events containing $\pi^0$ mesons, the $\pi^0$ candidates are selected by minimizing $|M_{\gamma\gamma} - M_{\pi^0}|$, where $M_{\pi^0}$ is the nominal $\pi^0$ mass from the PDG \cite{pdg1}.~A requirement of $|M_{\gamma \gamma}- M_{\pi^0}| > 20 \, \text{MeV/$c$}^2$ is applied, corresponding to approximately three times the mass resolution.~For the process $\psi(3686)\to\Sigma^{0}\bar{\Sigma}^{0}\eta$, to reject backgrounds originating from $\psi(3686)\to\eta J/\psi$ with the subsequent decay $J/\psi\to\Sigma^{0}\bar{\Sigma}^{0}$, the requirement $|M_{\Sigma^{0}\bar{\Sigma}^{0}} - M_{\jpsi}| > 30\, \text{MeV/$c$}^2$ is imposed, where $M_{\jpsi}$ denotes the nominal $\jpsi$ mass taken from the PDG \cite{pdg1}.~For $J/\psi\to\Sigma^{0}\bar{\Sigma}^{0}\eta$, to suppress the backgrounds with one fewer photon, an additional 4C kinematic fit under the hypothesis of $J/\psi \to\Lambda \bar{\Lambda}\gamma\gamma\gamma$ is performed for all possible photon combinations. Events with $\chi^ {2}_{4\rm{C}}$ being less than that of the signal hypothesis are discarded.~The combination with the minimum $\chi^{2}_{\gamma\gamma\gamma}$ is selected and compared with the $\chi^{2}_{4\rm{c}}$ from the signal hypothesis. Events with $\chi^{2}_{\gamma\gamma\gamma} < \chi^{2}_{4\rm{C}}$ are discarded.

\begin{figure*}[t]
 \centering
 \includegraphics[width=0.32\linewidth]{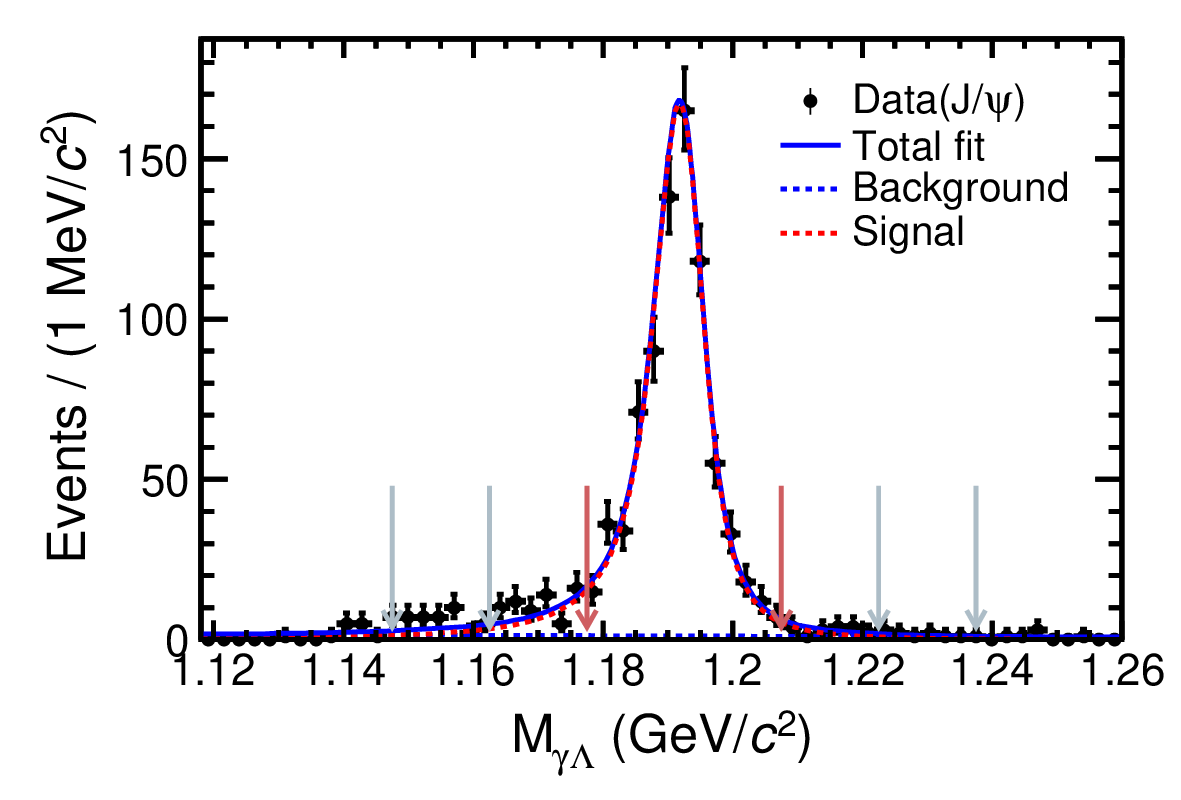}
 \includegraphics[width=0.32\linewidth]{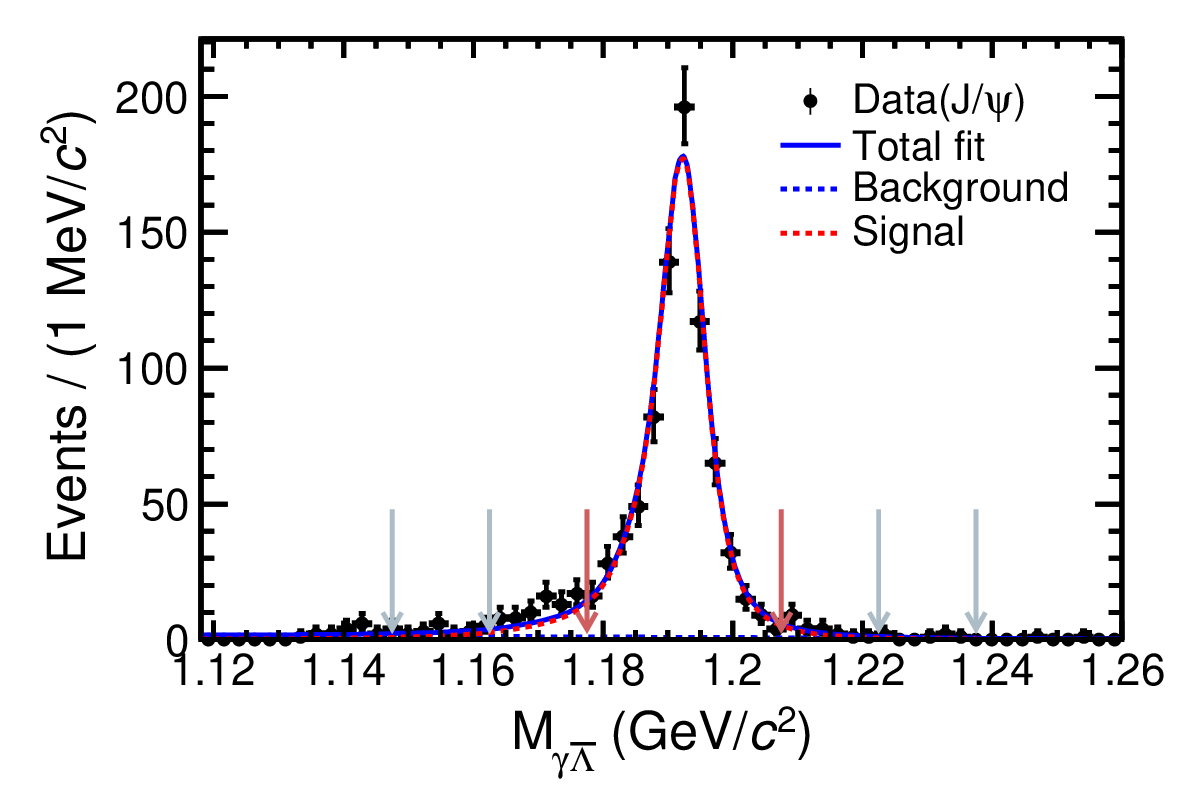}
 \includegraphics[width=0.32\linewidth]{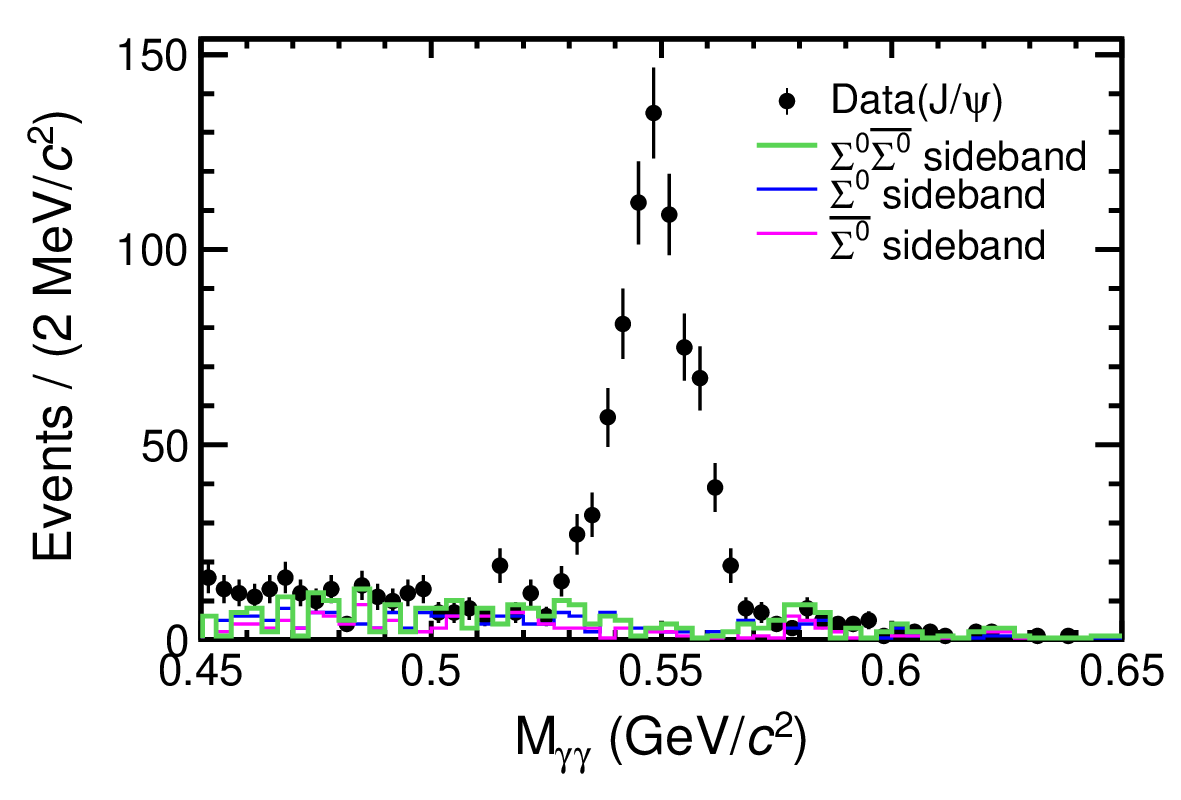}
 \put(-130,88){ (c)}

 \includegraphics[width=0.32\linewidth]{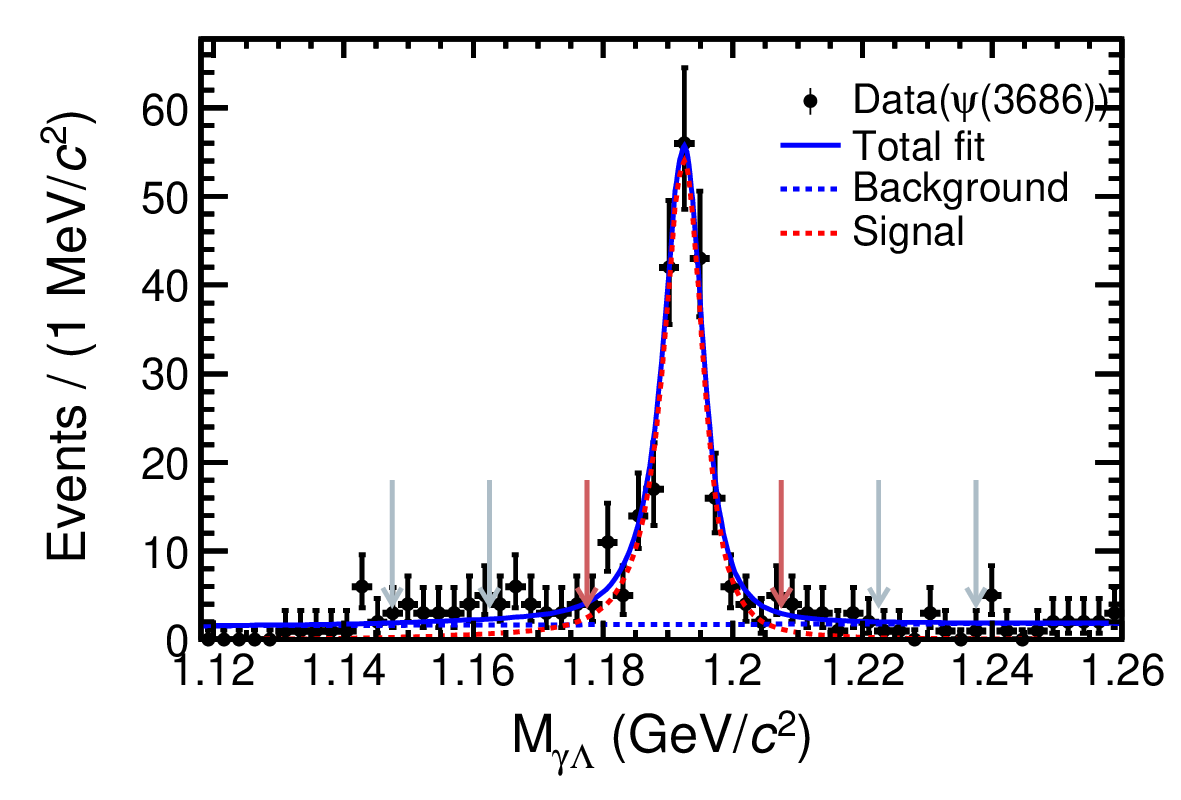}
 \includegraphics[width=0.32\linewidth]{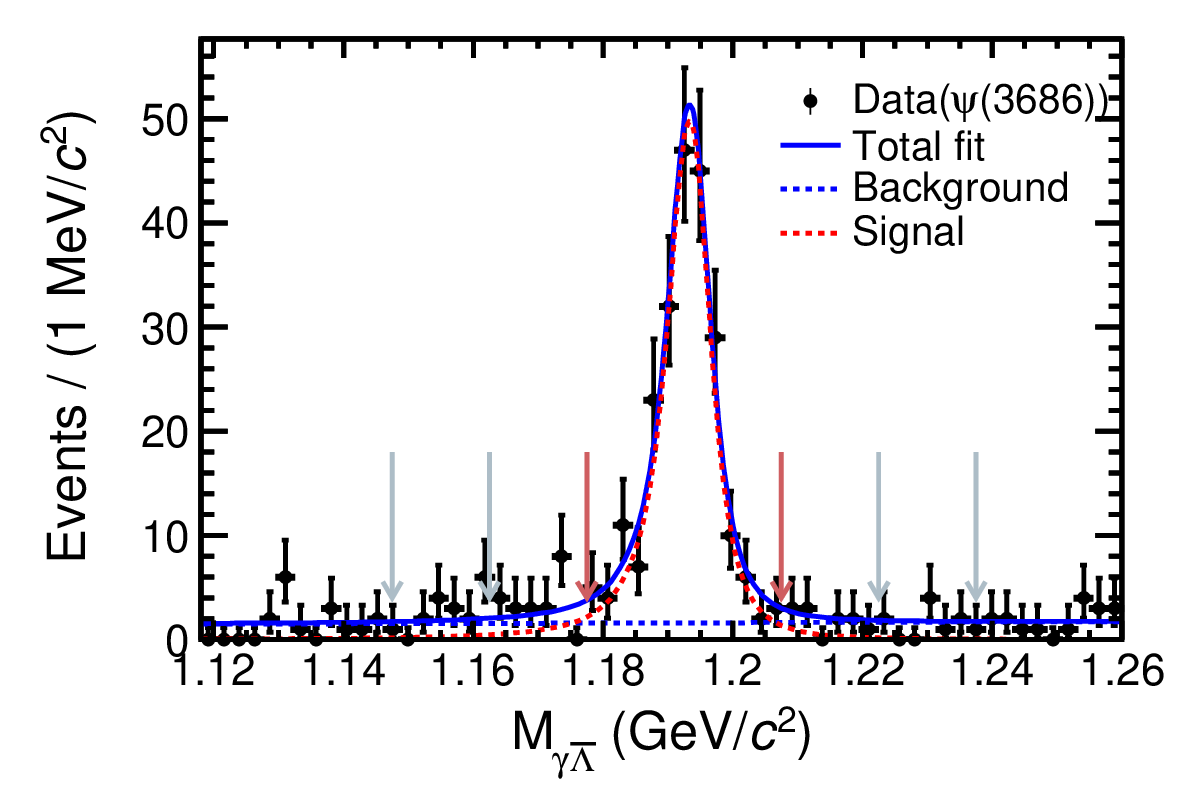}
 \includegraphics[width=0.32\linewidth]{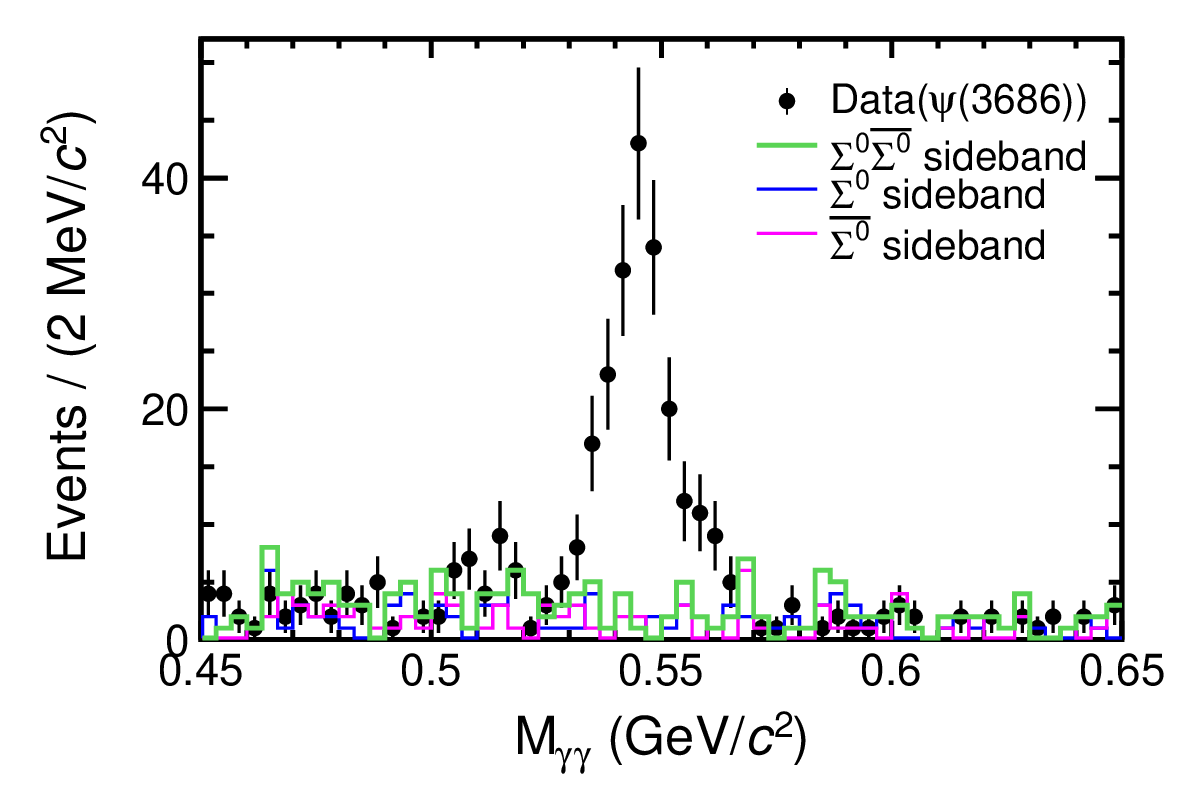}
 \put(-130,88){ (f)}
  \vspace*{-0.5cm}
\setlength{\abovecaptionskip}{1pt}
\caption{Invariant mass distributions of $\gamma\Lambda$, $\gamma\bar{\Lambda}$, $\gamma\gamma$ for $J/\psi\to\Sigma^{0}\bar{\Sigma}^{0}\eta$ (top) and $\psi(3686)\to\Sigma^{0}\bar{\Sigma}^{0}\eta$ (bottom). The dots with error bars are the data, the blue solid lines give the total fit results, the red dashed lines are the signal components, and the blue dashed lines are the smooth backgrounds. The red and gray solid arrows show the $\Sigma^0$ or $\bar \Sigma^0$ signal and sideband regions, respectively.}
 \label{charged:fit_sigma_2}
\end{figure*}

Potential backgrounds are studied using the inclusive MC samples, which include 10 billion $J/\psi$ events and 2.7 billion $\psi(3686)$ events.~No significant peaking background is observed in the invariant mass distribution of the $\gamma\gamma$.~Therefore, the two-dimensional $\Sigma^{0}\bar \Sigma^{0}$ sideband method is used to estimate combinatorial backgrounds by combining background events in the $\Sigma^{0}$ and $\bar{\Sigma}^{0}$ sideband regions:~[1.1475,~1.1625]~$\text{GeV/$c$}^2$ and [1.2225,~1.2375]~$\text{GeV/$c$}^2$.~The scale factor is determined by fitting the $\Sigma^0$ and $\bar{\Sigma}^{0}$ mass spectra, where the signal shape is described by the signal MC shape convolved with a Gaussian resolution function, while the background shape is described by a linear function.~The fit results are shown in Fig.~\ref{charged:fit_sigma_2}.~The  distributions of $M_{\gamma\gamma}$ from the $\Sigma^{0}$-$\bar{\Sigma}^{0}$ two-dimensional sideband region are shown in  Fig.~\ref{charged:fit_sigma_2} (c) and (f).

To estimate the background yield from continuum light hadron production,~which originates directly from $e^+ e^-$ annihilation rather than $\psi(3686)$ decays, we perform the same analysis procedure on the data samples at center-of-mass energies of 3.65 GeV, 3.682 GeV, and 3.773 GeV.~No event remains after applying all the selection requirements for the data taken at 3.65 and 3.682 GeV, and only one event remains at 3.773 GeV.~Due to limited statistics and the absence of structures significantly different from statistical fluctuations, the background from continuum light hadron production is neglected.
\section{Branching-fraction determination}
\begin{figure}[htbp]
  \centering
  \vskip 0.0cm
  \hskip -0.0cm \mbox{
  \begin{overpic}[width=7.0cm,height=5.0cm,angle=0]{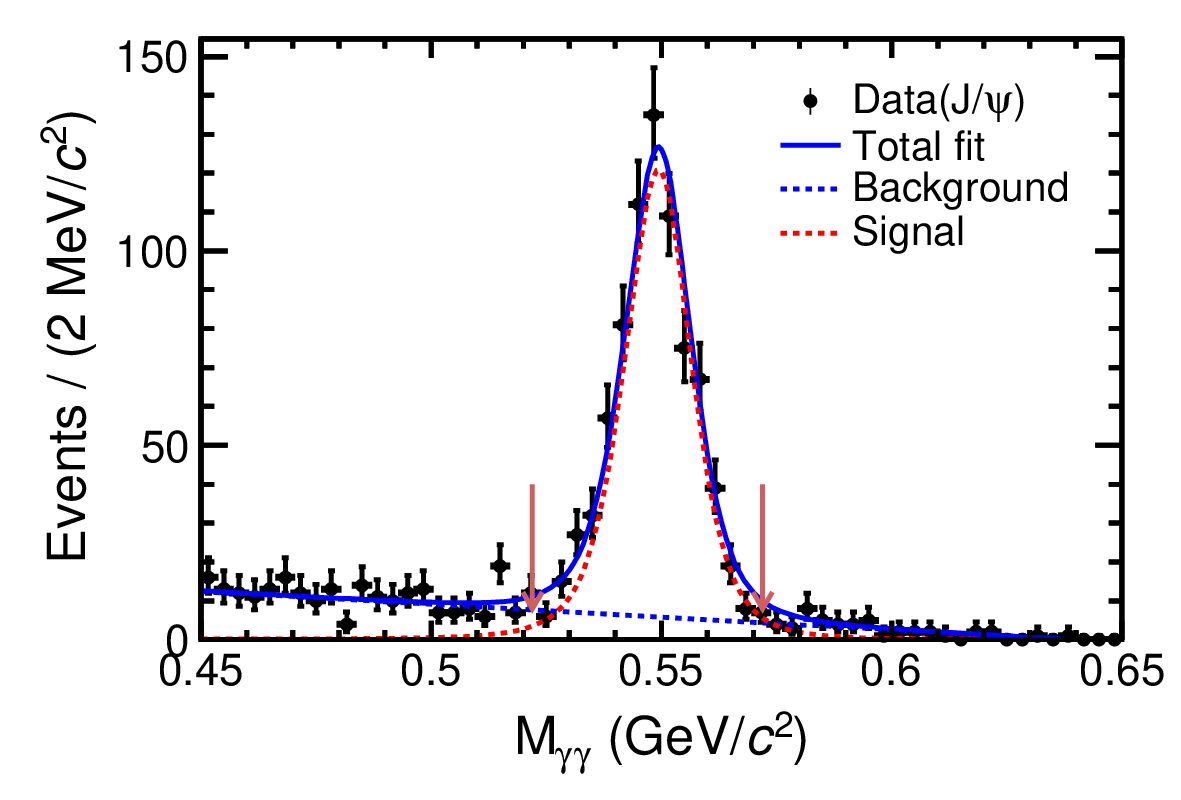}
   \put(22,58){(a)}
    \end{overpic}}
           \begin{overpic}[width=7.0cm,height=5.0cm,angle=0]{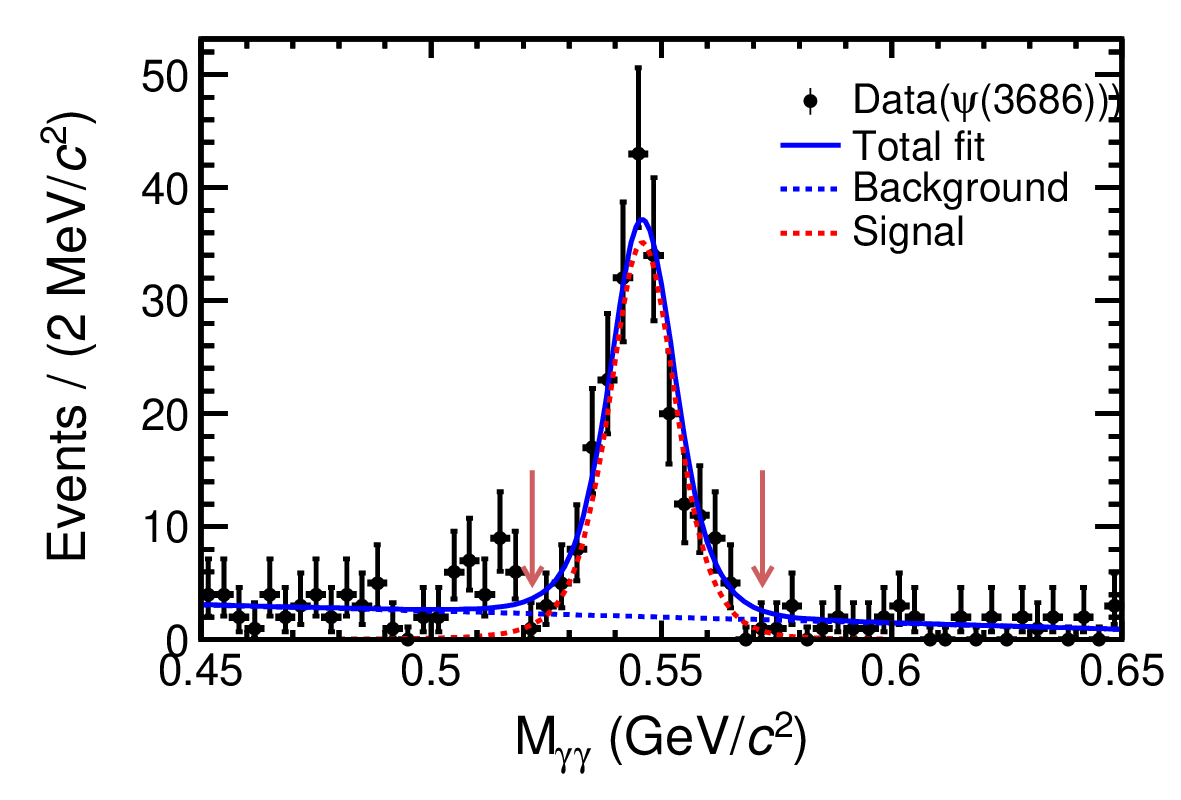}
           \put(22,58){(b)}
           \end{overpic}
   \caption{ Fits to the $M_{\gamma\gamma}$ distributions for the data. The top and bottom figures are the $\jpsi$ and $\psi(3686)$ samples, respectively. The dots with error bars show the data, the blue solid lines give the total fit results, the red dashed lines are the signal components, and the blue dashed lines are the smooth backgrounds. The region marked by the red arrows denotes the signal region.}
  \label{2body_egg}
        \end{figure}

To extract the signal yield, an unbinned maximum-likelihood fit is performed on the $M_{\gamma \gamma}$ distribution, where the signal is described by the MC-simulated shape convolved with a Gaussian function with free parameters in the fit to account for the resolution difference between data and MC simulation, and the background is described by a linear function.
Figure~\ref{2body_egg} shows the fitting results of $M_{\gamma \gamma}$, from which the signal yields are determined to be 193 $\pm$ 15 and 732 $\pm$ 30 for $\psi(3686) \to\Sigma^0\bar \Sigma^0\eta$ and $J/\psi \to\Sigma^0\bar \Sigma^0\eta$, respectively.

\begin{figure*}[htbp]
 \centering
 \includegraphics[width=0.32\linewidth]{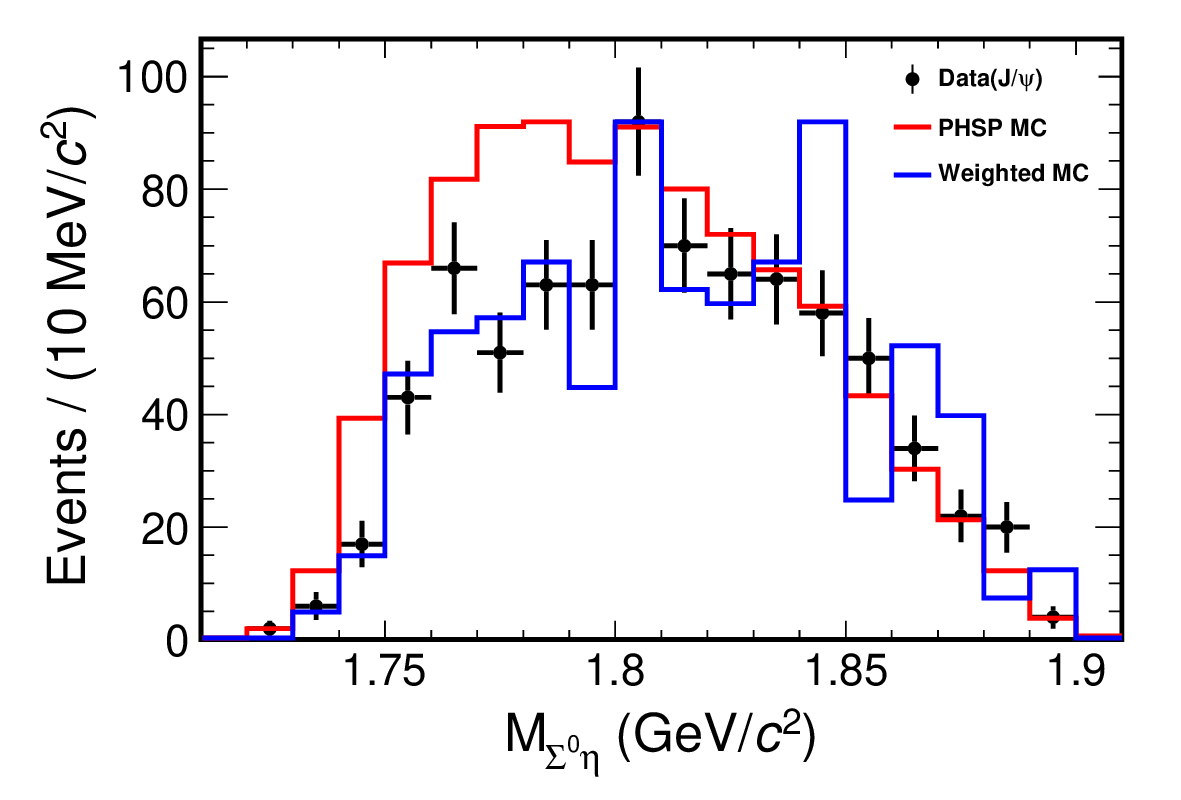}
 \put(20,57){}
 \includegraphics[width=0.32\linewidth]{jpsi-sigma0eta.eps}
 \put(20,57){}
 \includegraphics[width=0.32\linewidth]{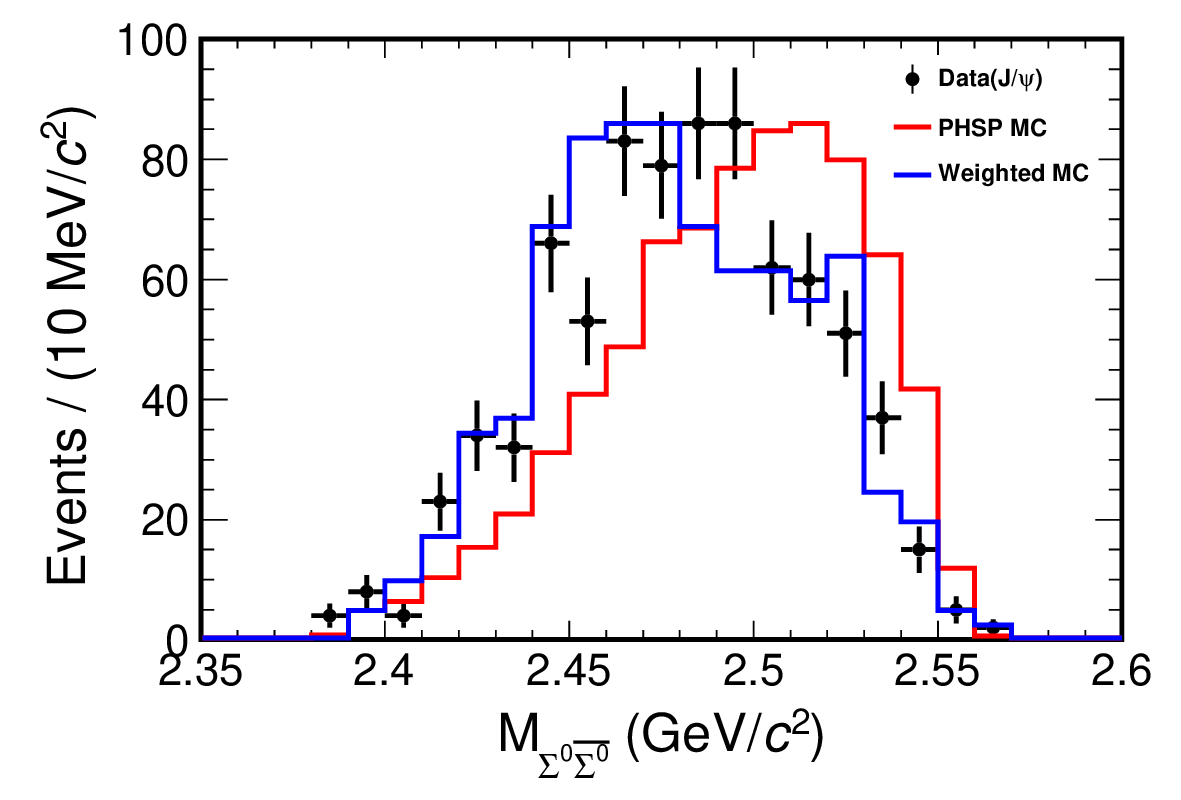}
 \put(20,57){}
 
 \includegraphics[width=0.32\linewidth]{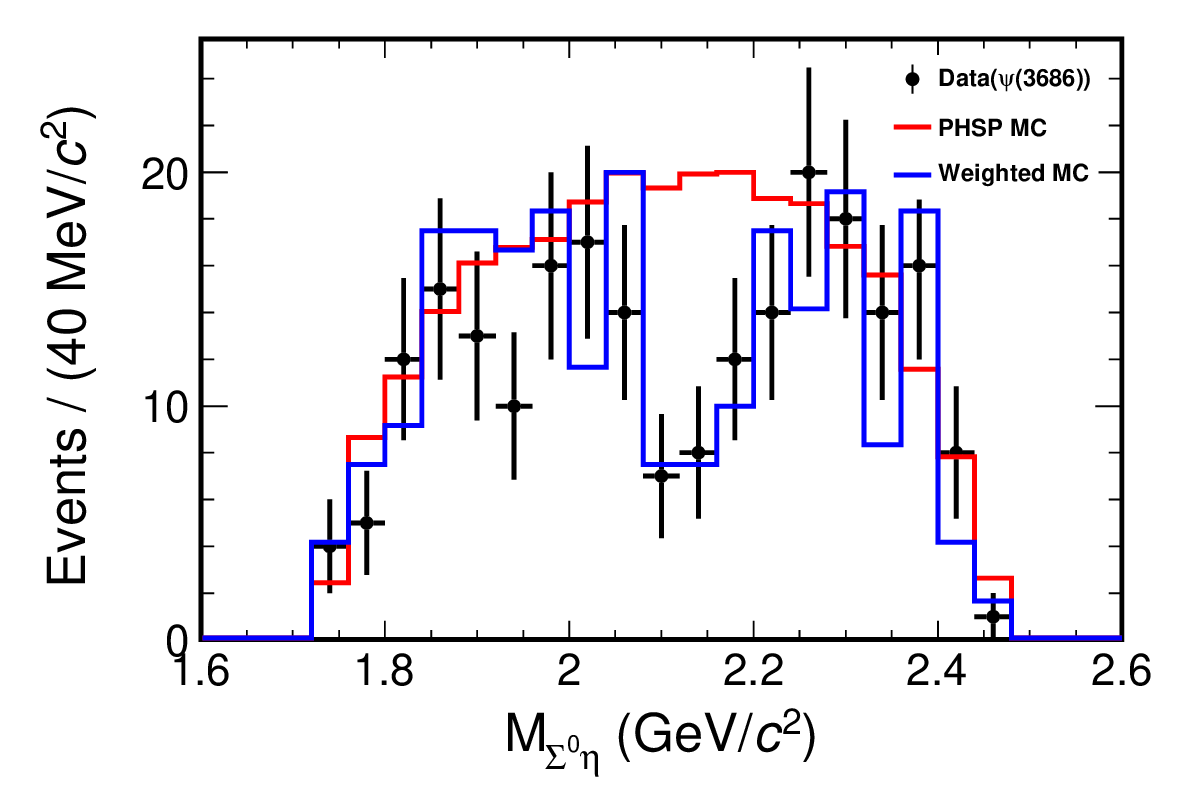}
 \put(20,57){}
 \includegraphics[width=0.32\linewidth]{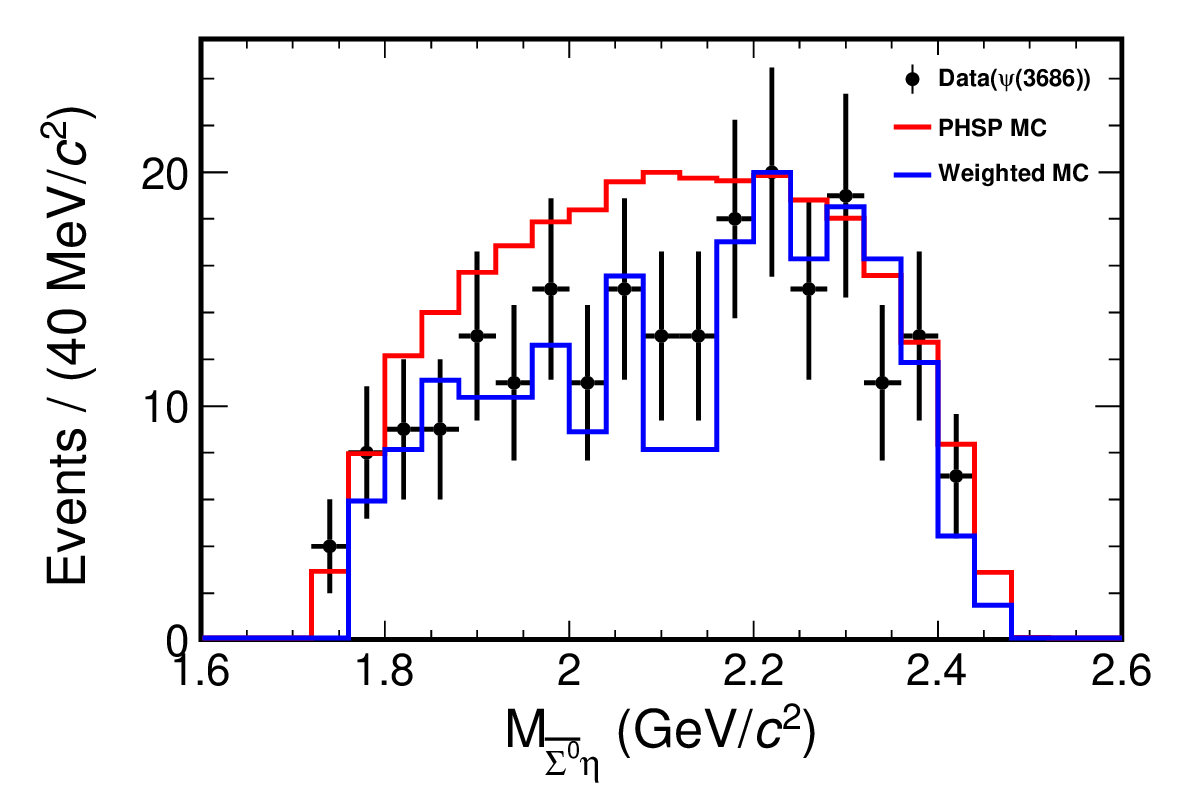}
 \put(20,57){}
 \includegraphics[width=0.32\linewidth]{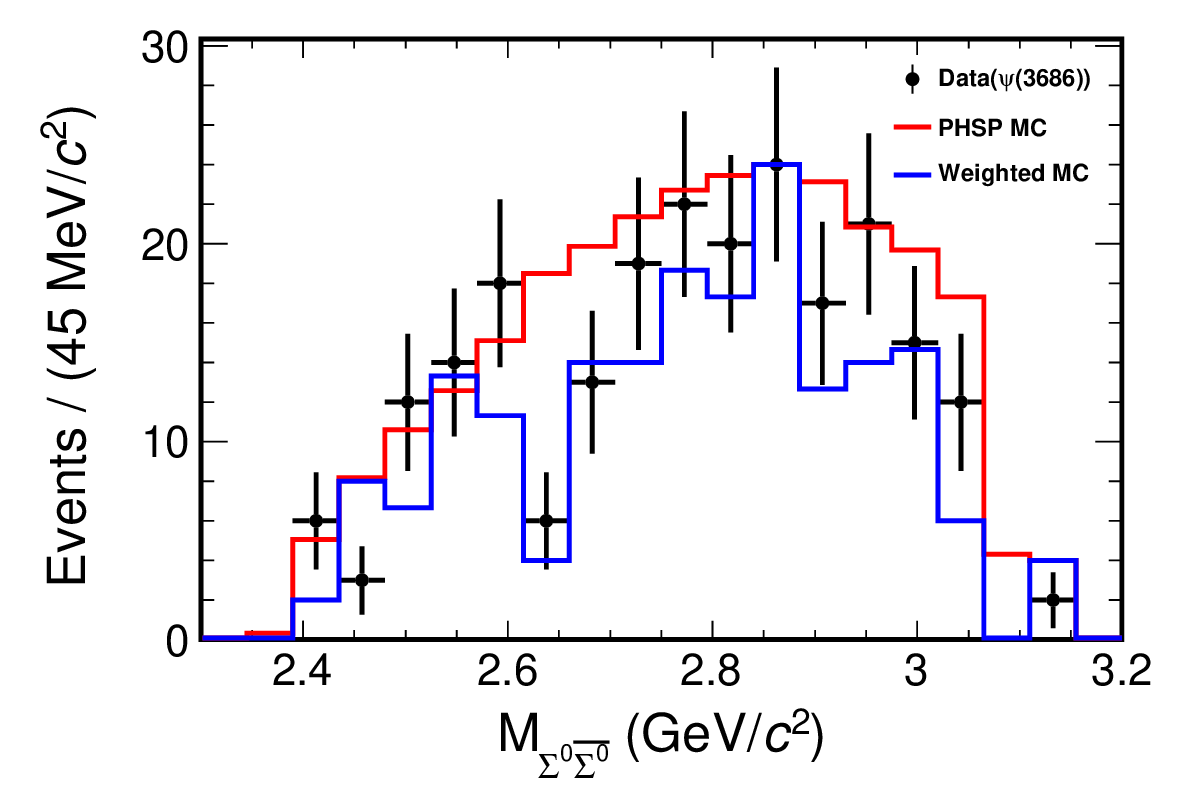}
 \put(20,57){}
  \vspace*{-0.5cm}
\setlength{\abovecaptionskip}{1pt}
\caption{Invariant mass distributions of $\Sigma^0\eta$, $\bar{\Sigma}^{0}\eta$ and $\Sigma^0\bar{\Sigma}^{0}$ for $J/\psi$ (top) and $\psi(3686)$ (bottom). The black dots with error bars represent data,
the red solid histograms represent the PHSP MC simulations, and the blue solid histograms represent the reweighted MC simulation.}
 \label{charged:w2}
\end{figure*}

Given the significant discrepancy between the data and the PHSP MC simulation (shown in Fig.~\ref{charged:w2}), the PHSP MC sample cannot provide an accurate estimate of the signal detection efficiency.~To reliably determine the detection efficiencies, an event-by-event weighting method is applied, in which PHSP MC events are reweighted according to the observed mass distribution. Specifically, the weight factor is derived from the two-dimensional ratio between the data and the PHSP MC in the $\Sigma^0\eta$ versus $\bar{\Sigma^0}\eta$ Dalitz plot, and is used to correct the signal MC sample. The specific formula is as follows:
\begin{equation}
\varepsilon_{w} \equiv \frac{\sum_{i=1}^{N^{\text{selected}}} w_{i}}{\sum_{i=1}^{N^{\text{generated}}} w_{i}}.
\end{equation}
Here, $\omega_i$ represents the weight factor for the event $i$ of the PHSP MC sample, while $N^{\mathrm{selected}}$ and $N^{\mathrm{generated}}$ denote the numbers of selected and generated events, respectively.~The final efficiencies are $0.6\%$ and 3.4\% for the $J/\psi$ and $\psi(3686)$ decays, respectively.
The branching fraction of the $\psi(3686)\to \Sigma^0\bar \Sigma^0\eta$ decay is calculated by
\begin{equation}
 \mathcal{B}(\psi'\to \Sigma^0\bar{\Sigma}^{0}\eta)\equiv\frac{N_{\mathrm{sig}}}{N_{\psi'}\cdot\varepsilon_{w}\cdot\mathcal{B}_{1}^{2}\cdot\mathcal{B}_{2}^{2}\cdot\mathcal{B}_{3}},
 \end{equation}
 where $N_{\psi'}$ is the total number of $\psi(3686)$ or $J/\psi$ events in the data; and $\mathcal{B}_{1}$,  $\mathcal{B}_{2}$, and $\mathcal{B}_{3}$ 
 are the branching fractions of~$\Sigma^0\to\gamma\Lambda$, $\Lambda\to p \pi^{-}$ and $\eta\to\gamma\gamma$ quoted from the PDG~\cite{pdg1}, respectively.~Finally, we obtain $\mathcal{B} (\psi(3686)\to\Sigma^0\bar{\Sigma}^{0}\eta)$ = $(1.3 \pm  0.1)\times 10^{-5}$ and $\mathcal{B} (J/\psi\to\Sigma^0\bar{\Sigma}^{0}\eta)$ = $(7.5 \pm  0.3)\times 10^{-5}$, where the uncertainties are statistical only.
 
No evidence is observed for excited $\Sigma^*$ states or threshold enhancements in the $\Sigma^0$($\bar{\Sigma}^{0}$)$\eta$ and $\Sigma^0$$\bar{\Sigma}^{0}$ invariant mass spectra.

\section{systematic uncertainty}
The sources of systematic uncertainties are summarized as follows.~The uncertainties in the measurements of the total numbers of $J/\psi$ and $\psi(3686)$ events in data are 0.4$\%$ \cite{ref_jpsinumber} and 0.5$\%$ \cite{ref_psinumber}, respectively, and the uncertainty of the photon detection is $0.5\%$ \cite{gamma} per photon. The systematic uncertainty due to the $\Lambda/\bar{\Lambda}$ reconstruction is estimated to be 2$\%$ \cite{222}. 

To estimate the systematic uncertainty arising from the 4C kinematic fit, corrections to the helix parameters of the MC simulation are applied to better align the MC with the data \cite{ref_M. Ablikim2013}.~The difference in signal efficiencies between the corrected and uncorrected cases is 0.1$\%$ for $J/\psi\to\Sigma^0\bar{\Sigma}^{0}\eta$ and is neglected for $\psi(3686)\to\Sigma^0\bar{\Sigma}^{0}\eta$.

The branching fractions of the sub-resonances are quoted from the PDG \cite{pdg1}, with their uncertainties of 0.8$\%$, 0.8$\%$ and 0.5$\%$ for $\mathcal{B}(\Lambda\to p \pi^{-})$, $\mathcal{B}(\bar{\Lambda} \to \bar{p}\pi^+)$ and $\mathcal{B}(\eta\to\gamma\gamma)$, respectively. The total uncertainty from the quoted branching fractions is 1.7$\%$ obtained by adding the individual contributions in quadrature.

To estimate the uncertainty of the MC model, the 3-body MC generator is used, with efficiency-corrected Dalitz plots of the $ M^2_{\Sigma^0 \eta} $ and $ M^2_{\bar{\Sigma}^{0} \eta} $ distributions as inputs.~Background events in data are subtracted using the two-dimensional sideband method.~To investigate the systematic uncertainties introduced by efficiency calculations using the two-dimensional weighting method, a new background-subtracted MC sample is generated, where the number of background events is fixed to the value determined by the background events counted in the $\Sigma^0\bar{\Sigma}^{0}$ two-dimensional mass from the final fit results.~Then, the input Dalitz plot is updated to the new background-subtracted data when simulating the signal decay process.~The changes in efficiencies are assigned as the systematic uncertainties, which are 7.1$\%$ for $J/\psi\to\Sigma^0\bar{\Sigma}^{0}\eta$ and 0.3$\%$ for $\psi(3686)\to\Sigma^0\bar{\Sigma}^{0}\eta$. 

The systematic uncertainties from the mass resolution are estimated by fitting the mass spectrum with the MC shape convolved with a fixed-parameter Gaussian function, where the fixed parameters are extracted from the control sample of $J/\psi \to \Sigma^{0}\bar{\Sigma}^{0}$.~The changes in detection efficiencies, $2.3\%$ for $J/\psi\to\Sigma^0\bar{\Sigma}^{0}\eta$ and $1.4\%$ for $\psi(3686)\to\Sigma^0\bar{\Sigma}^{0}\eta$, are assigned as the corresponding systematic uncertainties.

The~systematic uncertainties from the fitting procedure include the fit range,~the signal shape,~and the background shape.~To estimate the systematic uncertainties from the fit range,~several alternative fits are performed over different intervals: ([0.44,~0.64]~GeV/$c^{2}$,~[0.46,~0.66]~GeV/$c^{2}$,~[0.44,~0.66]~GeV/$c^{2}$, [0.46,~0.64]~GeV/$c^{2}$).~The largest resulting difference in signal yield is taken as the corresponding systematic uncertainty.~To estimate the uncertainty due to the signal shape, the signal MC shape convolved with a Gaussian function with free parameters is used to fit the $\eta$ and $\Sigma$ mass distributions in the data.~The difference in efficiencies before and after smearing the Gaussian resolution function is taken as the systematic uncertainty.~To estimate the uncertainties from the smooth background shape, the fitting function is changed from a first-order Chebyshev polynomial to a third-order one.~The difference in the fitted signal yields between the nominal and the alternative background shapes is taken as the corresponding systematic uncertainty.~The systematic uncertainties of the fitting procedure are $5.5\%$ for $J/\psi\to\Sigma^0\bar{\Sigma}^{0}\eta$ and $2.6\%$ for $\psi(3686)\to\Sigma^0\bar{\Sigma}^{0}\eta$.

The total systematic uncertainties are found to be 10.0$\%$ for $\jpsi$ and 4.9$\%$ for $\psi(3686)$, after adding all contributions in quadrature. All systematic uncertainties are listed in Table~\ref{charged:systematic_errors}. 
\begin{table}[htbp]
	\caption{Relative systematic uncertainties in the branching fraction measurements, where the items marked with $*$ are common for both decays.}
	\label{charged:systematic_errors}
	\centering
	\begin{small}   \begin{tabular}{lcc}
		\hline
		\hline
		  Source 
&$J/\psi~(\%)$&$\psi(3686)~(\%)$\\
		\hline
		  $N_\psi'$&0.4&0.5
\\
		  Photon detection*&2.0&2.0\\
		  $\Lambda\bar{\Lambda}$ reconstruction*& 2.8&2.8\\
		  Kinematic fit*&0.1&0.0\\
 Quoted branching fractions*& 1.7&1.7\\
		  MC model&7.1&0.3\\
		  Mass window*&2.3&1.4\\
		  Fit procedure&5.5&2.6\\
		\hline
		  Total&10.0&4.9\\
		\hline
		\hline
	\end{tabular}
\end{small}
\end{table}

\section{summary} 
Using $(10087 \pm 44) \times 10^6$ $J/\psi$ and $(2712 \pm 14) \times 10^6$  $\psi(3686)$ events collected with the BESIII detector, the decays $J/\psi \to \Sigma^0 \bar{\Sigma}^0\eta$ and $\psi(3686) \to \Sigma^0 \bar{\Sigma}^0\eta$ are observed for the first time, and the corresponding branching fractions are measured to be $\mathcal{B}(J/\psi \to \Sigma^0\bar{\Sigma}^0\eta)= (7.5 \pm 0.3\pm 0.8) \times 10^{-5}$ and $\mathcal{B}(\psi(3686) \to \Sigma^0\bar{\Sigma}^0\eta)= (1.3\pm 0.1 \pm 0.1) \times 10^{-5}$, where the first uncertainties are statistical and the second systematic.~The Q value is determined to be $(17.3 \pm 1.5 \pm 1.7)\%$, which is consistent with the 12\%-rule within 3.0$\sigma$.~Here, the common systematic uncertainties of tracking, PID, photon reconstruction, mass window and cited branching fractions have been canceled.~In addition, no significant structures are observed in the $\Sigma^0$($\bar{\Sigma}^0$)$\eta$ and $\Sigma^0$$\bar{\Sigma}^0$ invariant mass spectra.

\section{Acknowledgement}

The BESIII Collaboration thanks the staff of BEPCII (https://cstr.cn/31109.02.BEPC) and the IHEP computing center for their strong support. This work is supported in part by National Key R\&D Program of China under Contracts Nos. 2025YFA1613900, 2023YFA1606000, 2023YFA1606704; National Natural Science Foundation of China (NSFC) under Contracts Nos. 11635010, 11935015, 11935016, 11935018, 12025502, 12035009, 12035013, 12061131003, 12192260, 12192261, 12192262, 12192263, 12192264, 12192265, 12221005, 12225509, 12235017, 12342502, 12361141819; the Chinese Academy of Sciences (CAS) Large-Scale Scientific Facility Program; the Strategic Priority Research Program of Chinese Academy of Sciences under Contract No. XDA0480600; CAS under Contract No. YSBR-101; 100 Talents Program of CAS; The Institute of Nuclear and Particle Physics (INPAC) and Shanghai Key Laboratory for Particle Physics and Cosmology; ERC under Contract No. 758462; German Research Foundation DFG under Contract No. FOR5327; Istituto Nazionale di Fisica Nucleare, Italy; Knut and Alice Wallenberg Foundation under Contracts Nos. 2021.0174, 2021.0299, 2023.0315; Ministry of Development of Turkey under Contract No. DPT2006K-120470; National Research Foundation of Korea under Contract No. NRF-2022R1A2C1092335; National Science and Technology fund of Mongolia; Polish National Science Centre under Contract No. 2024/53/B/ST2/00975; STFC (United Kingdom); Swedish Research Council under Contract No. 2019.04595; U. S. Department of Energy under Contract No. DE-FG02-05ER41374


 

\end{document}